\documentclass{ifacconf}

\usepackage{graphicx}      
\usepackage{natbib}        
\usepackage{amssymb,amsmath}
\usepackage{csquotes}
\usepackage{mathrsfs}

\usepackage{subfigure}
\newtheorem{theorem}{Theorem}{}
{}
{}
\newtheorem{remark}{Remark}{}
\begin{document}
\begin{frontmatter}

\title{LPV Delay-Dependent Sampled-Data Output-Feedback Control of Fueling in Spark Ignition Engines} 


\author{Shahin Tasoujian,} 
\author{Karolos Grigoriadis,} 
\author{Matthew Franchek}

\address{Department of Mechanical Engineering, University of Houston, Houston, TX, 77004 USA (e-mail: stasoujian@uh.edu).}

\begin{abstract}                
We propose a delay-dependent sampled-data output-feedback LPV control technique to address the air-fuel ratio (AFR) regulation problem in spark ignition (SI) engines. AFR control and advanced fueling strategies are essential for maximizing the fuel economy while minimizing harmful exhaust emissions. The fuel path of the SI engine, as well as the three-way catalyst (TWC) simplified dynamics, have been captured by a continuous-time linear parameter-varying (LPV) system with varying time delay, where the system dynamics rely on the engine speed, defined as the system's scheduling parameter. The interconnection of the continuous-time plant and a digital controller through analog-to-digital and digital-to-analog converter devices forms a hybrid closed-loop configuration. Therefore, in order to benefit from continuous-time control synthesis tools, the input-delay method has been employed to transform the hybrid closed-loop system into the continuous-time domain with system inherent time delay and an additional delay imposed by the mapping approach. The designed sampled-data gain scheduled output-feedback controller seeks to establish the closed-loop asymptotic stability and a prescribed level of performance for the LPV system with an arbitrarily varying time delay and varying sampling time, where the synthesis results are provided in a convex linear matrix inequality (LMI) constraint setting. Finally, several closed-loop simulation scenarios are conducted, and comparisons are provided to demonstrate the proposed methodology's performance in achieving precise reference AFR tracking and disturbance attenuation.
\end{abstract}


\end{frontmatter}

\section{Introduction}
%
%
%
%
In the recent decades, due to strict environmental standards for the emission levels of exhaust gases from automobile engines and also economical demands, there has been considerable growing research and advancements in the field of air-fuel ration (AFR) control in the automotive engines. A three-way catalyst (TWC) is used in the exhaust after-treatment system to reduce harmful exhaust tail-pipe emissions, such as NOx, CO, and HC. The efficiency of TWC is closely correlated to the ratio of air mass to the mass of fuel inducted in the cylinder and in order to maintain its highest emission filtering efficiency, AFR should be maintained in a narrow stoichiometric region see \cite{guzzella2009introduction}. Therefore, efficient fueling strategies, \textit{i.e.} AFR control, play a crucial rule in increasing TWC efficiency, reducing fuel consumption, and enable torque control. The spark ignition (SI) engine AFR is regulated by manipulating the fuel injector pulse width, based on the feedback signal received from the Universal Exhaust Gas Oxygen (UEGO) sensor located upstream of the TWC. The SI engine AFR control problem has been widely addressed in the control literature and various control design methods has been proposed, such as loop-shaping control \cite{tasoujian2016}, PID control \cite{ebrahimi2012parameter}, and LPV-based fuzzy techniques \cite{wu2018air}.

A main challenge in the AFR control problem in SI engines stems from the variable nature of the engine operation due to varying engine speed which leads to a time-varying delay in the exhaust system feedback loop. The delay in the AFR control problem consists of a cycle delay and an exhaust gas transport delay, and the total time-delay refers to the time that it takes for the air-fuel mixture to reach the UEGO sensor downstream of the exhaust valve. The time-delay presence is introducing a phase lag and reduces the stability margin of the closed-loop control system, which can lead to performance deterioration or instability \cite{niculescu2012advances, Tasoujian2020thesis}. 

In the controls literature, two directions have been considered for the stability and performance analysis of time-delay systems, namely,  delay-independent and delay-dependent approaches. The delay-independent stabilization criterion results in conditions that are independent of the size of the delay. Besides its relatively effortless design process, the delay-independent approach suffers from conservative results, especially for systems with smaller time delays \cite{mahmoud1994design}. On the other hand, delay-dependent criteria result in less conservative analysis and guarantee the stability and the prescribed performance of the system by considering all the delays smaller than a certain determined upper bound. Time-domain methods based on Lyapunov-Krasovskii functionals and Lyapunov-Razumikhin functions have been utilized in \cite{fridman2014introduction} to examine the stability of linear time-invariant (LTI) time-delay systems. Additionally, the delay-dependent approach using Lyapunov-Krasovskii functionals has been also developed for the stability analysis and control synthesis of LPV time-delay systems (see \cite{loiseau2009topics, sakthivel2012robust}), where an LPV formulation is utilized to systematically characterize a nonlinear or a time-varying system through a class of linear systems whose dynamics depend on a measurable scheduling parameter \cite{briat2014linear}.

The literature on the continuous-time LPV time delay control system design is rich \cite{tasoujian2019delay, tasoujian2019robust, tasoujian2020robust, tasoujian2020scaled, tasoujian2021improved}. However, the controller implementation is typically accomplished by a digital device in discrete-time. This combination of the continuous-time plant and the discrete-time digital controller along with the sampling analog-to-digital (A/D) and the holding digital-to-analog (D/A) converter devices forms a sampled-data closed-loop system representation of hybrid nature \cite{ramezanifar2015sampled}. To this end, the designer is required to find an appropriate discrete-time controller to provide guarantees of stability and prescribed level of performance for the closed-loop hybrid system while taking the converter devices and intersample behavior into account \cite{chen2012optimal}. The sampled-data control design process is even more demanding when the studied continuous-time plant is a nonlinear or an LPV system \cite{tan2002output}. A general intuitive approach, also known as the indirect digital controller design, is to discretize the plant,  find a discrete-time controller, and then cascade the digital controller with the continuous-time plant along with the converter devices \cite{chen2012optimal}. Another indirect method is to take advantage of well-established continuous-time control design methods to design a controller, and then use conventional methods such as trapezoidal approximation \cite{apkarian1997discretization} to end up with the controller in the discrete-time domain \cite{ramezanifar2015sampled}. Despite the simplicity in the design process,  traditionally utilized indirect sampled-data control design approaches fail to guarantee stability and desired performance for the closed-loop hybrid dynamical system and also disregard the effect of sampling/hold rate of the converter devices in the design process. This may result in the degraded performance of the control design and potential instability \cite{ramezanifar2014output}. On the other hand, the authors in \cite{ramezanifar2015sampled, tan2002output} used the lifting technique \cite{bamieh1992general} to come up with direct sampled-data filter or control design methods for the LPV systems. In the lifting method, first, the continuous-time plant and the sample and hold converter devices are augmented, and then the augmented plant is mapped to an equivalent infinite dimensional discrete-time system representation followed by a digital control design process.  However, the sampled-data control design methods relying on the lifting approach are considered to be computationally complex and cumbersome \cite{suplin2007sampled}.

In this paper, an LPV system with time-varying state-delay representation is used to capture the time-varying AFR and TWC dynamics of an SI engine, in which the dynamics characteristics are formulated to depend on the engine speed defined as the LPV scheduling parameter. Next, by employing the input-delay method \cite{fridman2014introduction}, we are enabled to develop a direct sampled-data LPV control method for the continuous-time LPV plant. The input-delay approach reformulates the digital control law in a delayed continuous-time form and subsequently transforms the hybrid closed-loop LPV time-delay system to a continuous-time one while capturing the varying inter-sample behavior of the system. A proper parameter-dependent Lyapunov-Krasovskii functional (LKF) candidate is selected to design a sampled-data output-feedback LPV controller. The controller is required to provide stability and to minimize the worst-case disturbance amplification in terms of the prescribed energy-to-energy (induced  $\mathcal{L}_2$ norm) performance specification of the closed-loop hybrid LPV system with fast-varying time-delay and sampling rate. The delay-dependent synthesis conditions are formulated in a numerically tractable convex linear matrix inequality (LMI) optimization framework that can be readily solved using available interior-point optimization tools. A conventional indirect sampled-data controller, derived using the trapezoidal discretization approach, is also designed as a comparison control method.  Finally, this controller, along with the proposed sampled-data LPV control design method are evaluated in a simulation environment. The results and comparisons confirm the superiority and the effectiveness of the proposed LPV fueling control strategy in the sense of the commanded AFR reference tracking, disturbance rejection and noise attenuation.

The paper is organized as follows. The mathematical description of the AFR dynamics in the SI engine and its simplified LPV model are provided in Section \ref{sec:Problemformulation}. Section \ref{sec:Control} presents the stability and performance analysis of sampled-data LPV time-delay systems leading to the sampled-data output-feedback LPV control synthesis conditions.  Section \ref{sec:Results} outlines the AFR control design and simulation results and evaluates the performance of the proposed LPV controller in comparison with a conventional controller using a discretization approach. Final remarks and future research directions are provided in section \ref{sec:Conclusion}. Proof of the results are omitted for brevity.

\section{AFR Dynamics and Modeling}\label{sec:Problemformulation}

\subsection{Spark Ignition Engine with TWC}
Various control strategies have been proposed to regulate the catalyst oxygen level, and this challenge has been one of the driving motivations behind the development of precise AFR control. In the present paper, we take into account the simplified catalyst dynamics as well as the simplified AFR dynamics dictated by the time-varying exhaust transport delay.
The input of interest is the fuel injector pulse-width, and the normalized AFR, $\lambda_{up}$ is being measured by the UEGO sensor upstream of the TWC. In \cite{guzzella2009introduction, kiencke2000automotive}, complicated first principles-based models have been proposed which are not suitable for the control design purposes. Accordingly, it is desirable to use a simplified engine model that captures the predominant dynamic features of the UEGO sensor in response to the changes in the fuel injector pulse width.
 
The investigated air-fuel subsystem of the engine plant can be modeled as a speed-dependent induction to exhaust stroke delay of the four-stroke cycle cascaded with the transport delay in the exhaust system. The fuel is injected into the manifold and sucked into the cylinders at the phase-shifted time periods, leading to a step characteristic in the step responses. These steps can be approximated by a first-order lag element. The overall dynamics of the air-fuel path can be modeled as a series combination of a first-order lag and a delay element as given by \cite{kiencke2000automotive}
\begin{equation}
G(s)= \frac{\Delta \lambda_{up}}{\Delta F_{\lambda}} = \frac{1}{T s+1} e ^ {-s\tau},
\label{eq:plant}
\end{equation}
 

\noindent where $\Delta \lambda_{up} = \lambda_{up} -1$ is the measured AFR by the UEGO sensor (the \enquote{up} subscript refers to upstream of TWC), $F_{\lambda}$ is the input fuel injector pulse width multiplier, $\Delta F_{\lambda} = F_{\lambda} -1$ is the incremental fuel injector pulse width. In (\ref{eq:plant}) the model parameter $T$ is the time constant and $\tau$ is the time delay. It is noted that the model parameters $T$ and $\tau$ depend on the operating point of the engine characterized by the engine's speed in rpm, $\omega$, and the mass of airflow $\dot{m}_a$ in the engine. Accordingly, the time constant $T$, depends on the engine speed $\omega$, and can be approximated by {\large $\frac{2 (CYL - 1)}{\omega . CYL}$}, where $CYL$ is the number of engine cylinders \cite{kiencke2000automotive}. The overall time delay of the system consists of two parts, the cycle delay $\tau_c$, and the gas transport delay $\tau_g$, \textit{i.e.} $\tau = \tau_c + \tau_g$. The cycle delay $\tau_c$, is the combustion time delay from the opening of the inlet valve until the opening of the exhaust valve, which depends on the engine speed and approximated to {\large $\frac{90}{\omega}$}. The gas transport delay $\tau_g$, is the time that it takes for the cylinder exhaust gas to reach the tailpipe UEGO sensor. The gas transport delay is estimated by $\tau_g=\frac{\alpha}{\dot{m}_a}$ where $\dot{m}_a$ is the air mass flow and $\alpha$ is an engine dependent constant that can be identified through experimental data and estimation techniques \cite{zhang2007linear}. The gas transport delay typically varies between 20 and 500 ms.  Assuming $\tau_c \approx \tau_g$, and for our case, considering a six-cylinder Ford F150 engine, the model parameters become
\begin{equation}
T=\frac{100}{\omega}, \: \: \: \: \: \: \: \: \tau = \tau_c + \tau_g = \frac{180}{\omega}.
\label{eq:model parameters}
\end{equation}
\noindent Now, by substituting (\ref{eq:model parameters}) in (\ref{eq:plant}), the overall AFR dynamics of the SI engine's fuel path system is approximated by a first-order model with time-delay in the continuous-time domain as follows:
\begin{equation}
\begin{array}{rcl}
\dot{x}(t) & = & -\dfrac{\omega(t)}{100} x(t) + \dfrac{\omega(t)}{100} u(t-\dfrac{180}{\omega(t)}),\\[0.15cm]
y(t)& = & x(t) + d(t),
\end{array}
\label{SI-engine model 1}
\end{equation}
\noindent where $u(t)$ is the input AFR, $\Delta F_{\lambda}$, corresponding to the fuel pulse width multiplier, $y(t)$ is the measured output AFR upstream the TWC, $ \Delta \lambda_{up}$, and $d(t)$ is a disturbance acting on the output.

A TWC converter is considered after the UEGO sensor for exhaust emissions reduction purposes. Modeling and control of TWCs have been a topic of extensive research and have been investigated broadly in the literature. The detailed thermodynamics based models of the TWC is complicated for control design purposes \cite{kumar2014spatio}. In the present paper, the TWC dynamics is considered as a limited simple integrator which models the oxygen storage behavior of the TWC as follows \cite{zope2009air}:
\begin{equation}
\Delta m_{O_2} = \frac{1}{s} (m_{O_{2,{up}}}) \Delta \lambda_{up},
\label{eq:twc}
\end{equation}
\noindent where {\large $m_{O_{2,{up}}}$} represents the mass oxygen flow upstream of the TWC. In the present work {\large $m_{O_{2,{up}}}$} is assumed to be unity. According to (\ref{eq:twc}), the stored oxygen in the TWC is related to the measured output AFR upstream the TWC, $ \Delta \lambda_{up}$. 
\subsection{LPV Modeling of AFR Dynamics}
To transform the input delay system (\ref{SI-engine model 1}) into a state-delay LPV representation, we introduce a new artificial dynamic feedback control, $u_a(t)$, given by
\begin{equation}
	u(s)=\frac{\Omega}{s + \Lambda} u_a (s),
\end{equation}
\noindent where $\Omega$ and $\Lambda$ are positive scalars that are selected based on the bandwidth of the actuators. In this paper, considering the engine AFR regulation problem, both $\Omega$ and $\Lambda$ values have been selected to be $\Omega = \Lambda = 50$ \cite{zope2009air}. A new augmented state vector can be defined as $\mathbf{x}^{\text{T}} _a = [\begin{array}{cc} x & u\end{array}]^{\text{T}}$. Hence, using $x_a(t)$ and $u_a(t)$, the LPV state-space representation of the SI engine plant shown in (\ref{SI-engine model 1}) is converted to the following state-delayed form:
\begin{small}
\begin{equation}
\begin{array}{rcl}
 \dot{\mathbf{x}}_a (t) & = &\begin{bmatrix}
-\dfrac{\omega(t)}{100} & 0\\ 
0 &  -\Lambda
\end{bmatrix} \mathbf{x}_a (t) + \begin{bmatrix}
0 & \dfrac{\omega(t)}{100}\\ 
0 & 0
\end{bmatrix}\mathbf{x}_a (t-\dfrac{180}{\omega(t)})\\[0.3cm]
& + &  \begin{bmatrix}
0\\ 
\Omega
\end{bmatrix}u_a (t),\\[0.4cm]
y_a (t) & = & \begin{bmatrix}
1 & 0
\end{bmatrix} \mathbf{x}_a (t) + d(t). 
\end{array}
\label{engine-plant-twostates}
\end{equation}
\end{small}
The main objective of the present AFR control problem is to track the given AFR reference command, $\lambda_{r}$, which defined as the ratio of actual AFR to stoichiometric AFR, and also keep the storage level of oxygen in the TWC at the desired level, in the presence fuel purge disturbances. To this end, two additional state equations, $\dot{x}_3 (t) = e(t) = r(t)-y_a(t) = r(t) - (x_1(t) + d(t))$, with $r(t)$ being the reference command to be tracked, and $\dot{x}_4 (t) = x_3 (t) = \int e(t)$ are introduced to the state-space representation of the SI engine plant (\ref{engine-plant-twostates}). It should be noted that $x_4$ is necessary for minimizing the effect of disturbances on the oxygen level stored in the TWC. By defining the engine speed as a system scheduling parameter, \textit{i.e.} $\rho(t) = \omega(t)$, the final augmented LPV time-delay system is represented as follows:
\begin{small}
\begin{equation}
\begin{array}{l}
 \dot{\mathbf{x}}_a (t)  = \begin{bmatrix}
- \frac{\rho(t)}{100} & 0 & 0 & 0\\ 
0 & -\Lambda & 0 & 0\\ 
-1 & 0 & -\epsilon_1 & 0\\ 
0 & 0 & 1 & -\epsilon_2
\end{bmatrix} \mathbf{x}_a (t)\\[1cm]
 +   \begin{bmatrix}
0 & \frac{\rho(t)}{100} & 0 & 0\\ 
0 & 0 & 0 & 0\\ 
0 & 0 & 0 & 0\\ 
0 & 0 & 0 & 0
\end{bmatrix}\mathbf{x}_a (t-\tau(t)) \!+\! \begin{bmatrix}
0 & 0\\ 
0 & 0\\ 
1 & -1\\ 
0 & 0
\end{bmatrix} \mathbf{w}(t) \!+\! \begin{bmatrix}
0\\ 
\Omega\\ 
0\\ 
0
\end{bmatrix} u_a (t),\\[0.8cm]
y (t)  =  \begin{bmatrix}
0 & 0 & 1 & 0
\end{bmatrix} \mathbf{x}_a (t).
\end{array}
\label{engine_final_LPV}
\end{equation}
\end{small}

\noindent The exogenous disturbance vector is $\mathbf{w}(t) = [\begin{array}{cc} r(t) & d(t)\end{array}]^{\text{T}}$, and the small positive scalars $\epsilon_1$ and $\epsilon_2$ are introduced for numerical solvability purposes. 

 																					
\section{Control Design Process}\label{sec:Control}

\subsection{LPV Time-Delay Systems and Design Objectives}
\begin{figure}[t] 
\centering \includegraphics[width=3.in,height=1.8in]{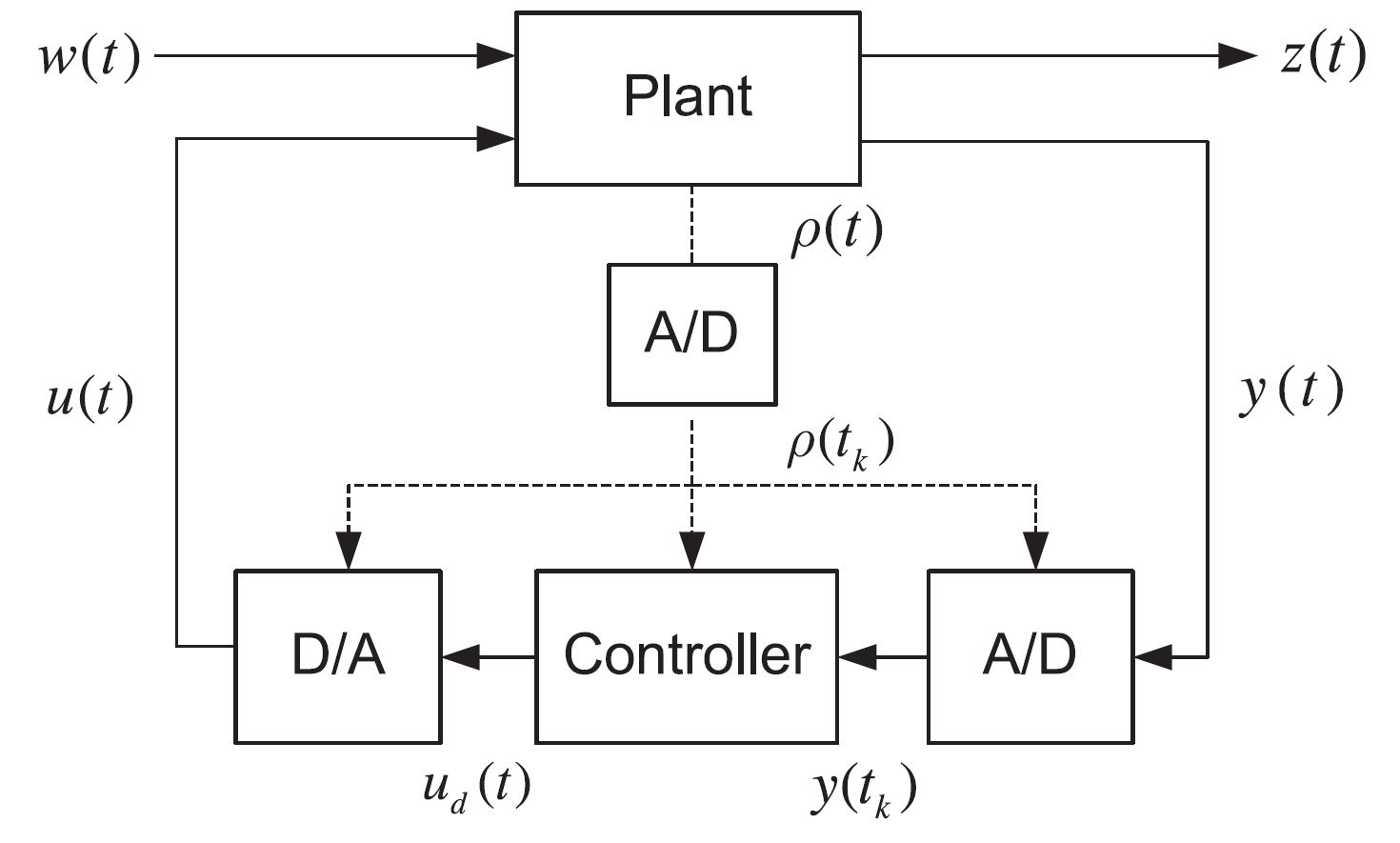} 
\caption{Sampled-data closed-loop system configuration} 
\label{fig:sampleddata_config}
\end{figure}

We consider the following state-space representation of a general LPV system with a varying state delay
\begin{equation}
\begin{array}{rcl}
 \dot{\mathbf{x}}(t) & =  & \mathbf{A}(\boldsymbol{\rho}(t)) \mathbf{x}(t)+\mathbf{A}_{\tau}(\boldsymbol{\rho}(t)) \mathbf{x}(t-\tau(t))\\[0.10cm] 
 & +  &   \mathbf{B}_1(\boldsymbol{\rho}(t))\mathbf{w}(t) + \mathbf{B}_2(\boldsymbol{\rho}(t))\mathbf{u}(t), \\[0.20cm] 
  \mathbf{z}(t)& = & \mathbf{C}_1(\boldsymbol{\rho}(t)) \mathbf{x}(t) + \mathbf{C}_{1 {\tau}}(\boldsymbol{\rho}(t)) \mathbf{x}(t-\tau(t)) \\[0.10cm]  
 & + &  \mathbf{D}_{11}(\boldsymbol{\rho}(t)) \mathbf{w}(t) + \mathbf{D}_{12}(\boldsymbol{\rho}(t)) \mathbf{u}(t),\\[0.20cm] 
 \mathbf{y}(t)& = & \mathbf{C}_2(\boldsymbol{\rho}(t)) \mathbf{x}(t), \\[0.20cm] 
 \mathbf{x}(t_0 + \theta)& = & \phi(\theta),\forall \theta \in [-\overline{\tau}, \: \: 0],
\end{array}
\label{eq:LPVsystem}
\end{equation}
where $\mathbf{x}(t) \in \mathbb{R}^n$ is the system state vector, $\mathbf{w}(t) \in \mathbb{R}^{n_w}$ is the vector of exogenous input with finite energy in the space $\mathcal{L}_2[0 \:\:\: \infty]$, $\mathbf{u}(t) \in \mathbb{R}^{n_u}$ is the control input vector, $\mathbf{z}(t) \in \mathbb{R}^{n_z}$ is the vector of outputs to be controlled, $\mathbf{y}(t) \in \mathbb{R}^{n_y}$ is the vector of measurable outputs. The system state space matrices $\mathbf{A}(\cdot)$, $\mathbf{A}_{\tau}(\cdot)$, $\mathbf{B}_1(\cdot)$, $\mathbf{B}_2(\cdot)$, $\mathbf{C}_1(\cdot)$, $\mathbf{C}_{1 {\tau}}(\cdot)$, $\mathbf{D}_{11}(\cdot)$, $\mathbf{D}_{12}(\cdot)$, $\mathbf{C}_2(\cdot)$ are assumed to be known continuous functions of the time-varying scheduling parameter vector $\boldsymbol{\rho}(\cdot) \in \mathscr{F}^\nu _\mathscr{P}$, where $\mathscr{F}^\nu _\mathscr{P}$ is the set of allowable parameter trajectories defined as
\begin{align}
 \mathscr{F}^\nu _\mathscr{P} \triangleq \{\boldsymbol{\rho}(t) \in \mathcal{C}(\mathbb{R}_{+},\mathbb{R}^{n_s}):\boldsymbol{\rho}(t) \in \mathscr{P}, |\dot{\rho}_i (t)| \leq \nu_i,\nonumber\\  i=1,2,\dots,n_s\},
 \label{eq:parametertraj}
\end{align}
where $n_s$ is the number of parameters and  $\mathscr{P}$ is a compact subset of $\mathbb{R}^{n_s}$. In (\ref{eq:LPVsystem}), $\boldsymbol{\phi}(\theta) \in \mathcal{C}([-\overline{\tau} \:\: 0], \mathbb{R}^n)$ is the functional system's initial condition, and $\tau (\boldsymbol{\rho}(t))$ is a differentiable scalar function representing the parameter-varying time delay which lies in the set $\mathscr{T}^{\nu_\tau}$ defined as  
\begin{align}
\mathscr{T}^{\nu_\tau} \triangleq\! \{ \tau (\boldsymbol{\rho}(t))\! \in \mathcal{C}(\mathscr{P},\mathbb{R}_{+}) \!:\! 0 \leq \tau (\cdot) \leq \overline{\tau} < \infty, \dot{\tau}(\cdot) \leq \nu_\tau\}.
\label{eq:delayset}
\end{align}
Due to the digital nature of controllers, in this paper, we seek to design a full-order discrete-time parameter-varying controller of the form
\begin{equation}
\begin{array}{rcl}
\mathbf{x}_d(k+1) & = & \mathbf{A}_d (\boldsymbol{\rho}(k)) \mathbf{x}_d(k)+ \sum_{i=1}^N \mathbf{A}_{\tau d_i} (\boldsymbol{\rho}(k)) \mathbf{x}_d(k-i)\\[0.10cm]
 & + & \mathbf{B}_d(\boldsymbol{\rho}(k))\mathbf{y}(k),\\[0.15cm]
\mathbf{u}_d(k) & = & \mathbf{C}_d (\boldsymbol{\rho}(k)) \mathbf{x}_d(k)+ \sum_{i=1}^N \mathbf{C}_{\tau d_i} (\boldsymbol{\rho}(k)) \mathbf{x}_d(k-i)\\[0.10cm]
& + & \mathbf{D}_d(\boldsymbol{\rho}(k))\mathbf{y}(k),
\end{array}
\label{eq:discretecontroller}
\end{equation}
\noindent that uses the sampled measurements of the continuous plant to generate a discrete control action. In (\ref{eq:discretecontroller}), $\mathbf{x}_K(k)$, $\mathbf{y}(k)$, and $\mathbf{u}_K(k)$ are the discrete-time instances of the controller state vector, measurement, and control input, respectively. For  sake of brevity, the index $k$ is chosen to show the sampling instances, $t_k$, and $N$ denotes the number of back samples of the controller state vector. Figure \ref{fig:sampleddata_config} demonstrates the configuration of the hybrid sampled-data closed-loop system showing the interconnection of the open-loop continuous-time system and a digital controller along with the signal converter devices.

Unlike a continuous-time LPV system in which the scheduling parameter vector $\boldsymbol{\rho}(t) \in \mathscr{F}^\nu _\mathscr{P}$ is continuously being measured in real-time, in the sampled-data control design framework, the parameter vector is only measured at the sampling instances. Hence, we assume that the parameter vector remains constant between two consecutive samples and the set of all admissible parameter trajectories is redefined for the sampled-data system as
\begin{multline}
\!\!\!\mathscr{\varepsilon}^\nu _\mathscr{P} \!\triangleq\! \{\boldsymbol{\rho}(t) \in \mathscr{P}, \boldsymbol{\rho}(t_k+t) = \boldsymbol{\rho}(t_k),|\rho_i(t_{k+1})-\rho_i(t_{k})| \leq \!\nu_i,\\
 k \in \mathbb{Z}_{+},i=1,2,\ldots,n_s, \forall t \in [0, \mathscr{T}_{k}] \},
\end{multline}
\noindent where $\mathscr{T}_{k}$ denotes the varying sampling period, \textit{i.e.} the length of time interval $[t_k, t_{k+1}]$. The interconnection of the system (\ref{eq:LPVsystem}), the controller (\ref{eq:discretecontroller}), and the converters form a closed-loop system $\mathbf{T}_{\mathbf{z}\mathbf{w}}$ which maps the disturbance signal $\mathbf{w}(t)$ to the desired control signal $\mathbf{z}(t)$. Considering the allowable parameter trajectories in $\mathscr{\varepsilon}^\nu _\mathscr{P}$, the designed controller is required to satisfy the following objective:

	\begin{itemize}
		\item Internal asymptotic stability  of the closed-loop system in the face of the parameters and delay variations, and disturbances to maintain the boundedness of the closed-loop system trajectories, and
		\item Minimization of the worst case amplification of the desired output, $\mathbf{z}$, to a nonzero disturbance signal, $\mathbf{w}$, with bounded energy, \textit{i.e.} solving the problem of $\gamma$-suboptimal induced $\mathcal{L}_2$-norm (energy-to-energy gain) of the mapping $\mathbf{T}_{\mathbf{z}\mathbf{w}}: \mathbf{w} \rightarrow \mathbf{z}$  given by
		
		\begin{equation}
    		{\min}\Vert \mathbf{T}_{\mathbf{z}\mathbf{w}}\Vert_{i,2} = {\min} \underset{\boldsymbol{\rho} \in \mathscr{\varepsilon}^\nu _\mathscr{P}}{\sup} \:\:\: \underset{\Vert \mathbf{w} \Vert_2 \neq 0 ,\mathbf{w} \in \mathcal{L}_2}{\sup}\:\: \frac{\Vert \mathbf{z} \Vert_2}{\Vert \mathbf{w} \Vert_2}<\gamma,
    		\label{eq:Performance Index}
		\end{equation}
	\end{itemize}
\noindent where $\gamma$ is a positive scalar.
\subsection{Stability and $\mathcal{H}_{\infty}$ Performance Analysis of Sampled-\\Data LPV Time-Delay Systems}

We consider a continuous-time full-order controller with a state-space representation as follows:
\begin{equation}
\begin{array}{rcl}
\dot{\mathbf{x}}_{K}(t) & = & \mathbf{A}_K (\boldsymbol{\rho}(t)) \mathbf{x}_K(t)+ \mathbf{A}_{\tau  K} (\boldsymbol{\rho}(t)) \mathbf{x}_K(t-\tau(\boldsymbol{\rho}(t)))\\[0.10cm]
 & + &\mathbf{A}_{\mathscr{T} K} (\boldsymbol{\rho}(t)) \mathbf{x}_K(t_k) + \mathbf{B}_K(\boldsymbol{\rho}(t))\mathbf{y}(t_k),\:\:\:\:\:\:\:\:\:\:\:\:\:\:\:\: \\[0.15cm]
\mathbf{u}_K(t_k) & = & \mathbf{C}_K (\boldsymbol{\rho}(t)) \mathbf{x}_K(t_k)+ \mathbf{D}_K(\boldsymbol{\rho}(t))\mathbf{y}(t_k),\:\:\:\:\:\:\:\:\:\:\:\:\:\:\:\:\:\:\:\:\\[0.10cm]
\mathbf{u} (t)& = & \mathbf{u}_K(t_k), \:\:\: t_k \leq t < t_{k+1},
\end{array}
\label{eq:contcont1}
\end{equation}
\noindent where $\boldsymbol{\rho} \in \mathscr{\varepsilon}^\nu _\mathscr{P}$ and $\boldsymbol{\rho}(t_k)$ is replaced with $\boldsymbol{\rho}(t)$ for $t_k \leq t < t_{k+1}$ for simplicity, emphasizing the fact that the continuous-time scheduling parameter vector is piecewise constant and does not vary in between sampling instances. It should be noted that unlike conventional approaches where a continuous-time controller is discretized without taking the converter devices into account, in the proposed sampled-data controller design approach (\ref{eq:contcont1}), the effects of sampling and holding devices are taken into consideration. In the sampled-data framework, as shown in Fig. \ref{fig:sampleddata_config}, controller (\ref{eq:contcont1}) uses discrete-time signals of the measurement and the scheduling parameter vector. In order to obtain a unified state-space continuous-time domain representation, the input-delay approach is utilized  \cite{fridman2014introduction}. We consider
\begin{equation}
u(t_k) = t (t - (t - t_k)) = u(t - \mathscr{T}_k(t)), \:\: t_k \leq t < t_{k+1},
\label{eq:inputdelayapproach}
\end{equation}
\noindent where $\mathscr{T}_k \leq t_{k+1} - t_k \leq \bar{\mathscr{T}}$ and $\bar{\mathscr{T}}$ denotes the maximum sampling interval value. Taking into account the input-delay approach, the sampling/holding characteristics are captured by defining a new delay term, $\mathscr{T}_k$. By substituting (\ref{eq:inputdelayapproach}) in (\ref{eq:contcont1}) the controller state-space model is transferred into a continuous-time representation as follows
\begin{small}
\begin{equation}
\begin{array}{lll}
\dot{\mathbf{x}}_{K}(t) \!&\! =\! &\! \mathbf{A}_K (\boldsymbol{\rho}(t)) \mathbf{x}_K(t)+ \mathbf{A}_{\tau  K} (\boldsymbol{\rho}(t)) \mathbf{x}_K(t-\tau(\boldsymbol{\rho}(t)))\\[0.1cm]
 \!&\! +\! &\!\mathbf{A}_{\mathscr{T} K} (\boldsymbol{\rho}(t)) \mathbf{x}_K(t - \mathscr{T}_k) \\[0.10cm]
 \!&\! +\! &\! \mathbf{B}_K(\boldsymbol{\rho}(t))\mathbf{C}_2(\boldsymbol{\rho}(t))\mathbf{x}(t \!-\! \mathscr{T}_k),\\[0.15cm]
\mathbf{u}(t) \!\!& \!\!=\! &\! \mathbf{C}_K (\boldsymbol{\rho}(t)) \mathbf{x}_K(t \!-\! \mathscr{T}_k)\!+\! \mathbf{D}_K(\boldsymbol{\rho}(t))\mathbf{C}_2(\boldsymbol{\rho}(t))\mathbf{x}(t \!-\! \mathscr{T}_k).
\end{array}
\label{eq:contcont2}
\end{equation}
\end{small}

\begin{theorem}\label{thm:thm1} There exists a full-order output-feedback gain-scheduled LPV controller (\ref{eq:contcont1}), over the sets $\mathscr{\varepsilon}^\nu _\mathscr{P}$ and $\mathscr{T}^{\nu_\tau}$, to asymptotically stabilizes the LPV system and satisfies the induced $\mathcal{L}_2$ norm performance level given in (\ref{eq:Performance Index}), with the sampling time $\mathscr{T}_k \leq \bar{\mathscr{T}}$, if there exists a continuously differentiable parameter dependent positive-definite  matrix function $\widetilde{\mathbf{P}}(\boldsymbol{\rho}(t)): \mathbb{R}^{s}\rightarrow\mathbb{S}^{2n}_{++}$, parameter dependent matrix functions $\mathbf{X}(\boldsymbol{\rho}(t)), \mathbf{Y}(\boldsymbol{\rho}(t)): \mathbb{R}^{s}\rightarrow\mathbb{S}^{n}$, positive-definite matrices $\widetilde{\mathbf{Q}}_\tau$, $\widetilde{\mathbf{Q}}_\mathscr{T}$, $\widetilde{\mathbf{R}}_{\tau}$, $\widetilde{\mathbf{R}}_{\mathscr{T}}$, $\widetilde{\mathbf{T}}_{\tau} \in \mathbb{S}^{2n}_{++}$, parameter dependent real matrix functions $\widehat{A}(\boldsymbol{\rho}(t))$, $\widehat{A}_{\tau}(\boldsymbol{\rho}(t))$, $\widehat{A}_{\mathscr{T}}(\boldsymbol{\rho}(t)): \mathbb{R}^{s}\rightarrow\mathbb{R}^{n \times n}$, $\widehat{B}(\boldsymbol{\rho}(t)): \mathbb{R}^{s}\rightarrow\mathbb{R}^{n \times n_u}$, $\widehat{C}(\boldsymbol{\rho}(t)): \mathbb{R}^{s}\rightarrow\mathbb{R}^{n_u \times n}$, $\widehat{C}_d (\boldsymbol{\rho}(t))$, $\mathbf{D}_K(\boldsymbol{\rho}(t)): \mathbb{R}^{s}\rightarrow\mathbb{R}^{n_u \times n_y}$, a positive scalar $\mathbf{\gamma}$, and real scalars $\lambda_2$, $\lambda_3$, $\lambda_4$, $\lambda_5 \in \mathbb{R}$ such that the LMI (\ref{eq:LMImain}) is feasible, where

\begin{figure*}[t]
\begin{equation}
\left[\begin{array}{ccccccc}
\widetilde{\boldsymbol{\Xi}}_{11} & \widetilde{\mathbf{P}}-\widetilde{\mathbf{V}} + \lambda_2\mathscr{A}^{\text{T}}  & \widetilde{\mathbf{R}}_\tau + \mathscr{A}_{\tau} +\lambda_3\mathscr{A}^{\text{T}} & \widetilde{\mathbf{R}}_{\mathscr{T}} + \mathscr{A}_{\mathscr{T}} +\lambda_4\mathscr{A}^{\text{T}}  & \lambda_5\mathscr{A}^{\text{T}} & \mathscr{B} & \mathscr{C}^{\text{T}} \\[2pt]
\star & \overline{\tau}^2 \widetilde{\mathbf{R}}_\tau + \overline{\mathscr{T}}^2 \widetilde{\mathbf{R}}_{\mathscr{T}} -2\lambda_2 \widetilde{\mathbf{V}} & \lambda_2 \mathscr{A}_{\tau} - \lambda_3 \widetilde{\mathbf{V}}  & \lambda_2 \mathscr{A}_{\mathscr{T}}  - \lambda_4 \widetilde{\mathbf{V}}   &  - \lambda_5 \widetilde{\mathbf{V}} & \lambda_2 \mathscr{B} & \mathbf{0}\\[2pt]
\star & \star & \widetilde{\boldsymbol{\Xi}}_{33} & \lambda_3 \mathscr{A}_{\mathscr{T}} + \lambda_4 \mathscr{A}^{\text{T}}_{\tau}  & \lambda_5 \mathscr{A}^{\text{T}}_{\tau}  & \lambda_3 \mathscr{B} & \mathscr{C}_{\tau}^{\text{T}} \\[2pt]
\star & \star & \star &  \widetilde{\boldsymbol{\Xi}}_{44} & \lambda_5 \mathscr{A}^{\text{T}}_{\mathscr{T}} &\lambda_4 \mathscr{B} & \mathscr{C}_{\mathscr{T}}^{\text{T}} \\[3pt]
\star & \star & \star & \star & -\widetilde{\mathbf{T}}_\tau & \lambda_5 \mathscr{B} & \mathbf{0}\\
\star & \star & \star  & \star & \star & -\gamma \mathbf{I}_{n_w} & \mathbf{D}_{11}^{\text{T}} \\
\star & \star & \star  & \star & \star  & \star & -\gamma \mathbf{I}_{n_z}
\end{array}\right]\prec\mathbf{0},
\label{eq:LMImain}
\end{equation}
\end{figure*}
\begin{small}
\begin{equation*}
\begin{array}{lll}
\widetilde{\mathbf{V}}  & = & \begin{bmatrix}
\mathbf{Y} & \mathbf{I}\\ 
\mathbf{I} & \mathbf{X}
\end{bmatrix},\\[0.3cm]
\mathscr{A} & = &\begin{bmatrix}
\mathbf{A}\mathbf{Y} & \mathbf{A}\\ 
\mathbf{X}\mathbf{A}\mathbf{Y}+ \mathbf{N}\mathbf{A}_K \mathbf{M}^{\text{T}} & \mathbf{X}\mathbf{A}
\end{bmatrix}=\begin{bmatrix}
\mathbf{A}\mathbf{Y} & \mathbf{A}\\ 
\widehat{A} & \mathbf{X}\mathbf{A}
\end{bmatrix},\\[0.4cm]
\mathscr{A}_{\tau} & = & \begin{bmatrix}
\mathbf{A}_{\tau}\mathbf{Y}& \mathbf{A}_{\tau}\\ 
\mathbf{X}\mathbf{A}_{\tau}\mathbf{Y} + \mathbf{N}\mathbf{A}_{\tau K} \mathbf{M}^{\text{T}} & \mathbf{X}\mathbf{A}_{\tau}
\end{bmatrix}=\begin{bmatrix}
\mathbf{A}_{\tau}\mathbf{Y}& \mathbf{A}_{\tau}\\ 
\widehat{A}_{\tau} & \mathbf{X}\mathbf{A}_{\tau}
\end{bmatrix},\\[0.4cm]
\mathscr{A}_{\mathscr{T}} \!& \!=\! &\!\!\!\! \begin{bmatrix}
\mathbf{B}_2(\mathbf{D}_K \mathbf{C}_2 \mathbf{Y} + \mathbf{C}_K \mathbf{M}^{\text{T}} )& \mathbf{B}_2 \mathbf{D}_K \mathbf{C}_2\\[0.3cm]
\begin{array}{cc}
\mathbf{X}\mathbf{B}_2\mathbf{D}_K\mathbf{C}_2\mathbf{Y} + \mathbf{N}\mathbf{B}_K\mathbf{C}_2\mathbf{Y}\\
+\mathbf{X}\mathbf{B}_2\mathbf{C}_K\mathbf{M}^{\text{T}} + \mathbf{N}\mathbf{A}_{\mathscr{T} K}\mathbf{M}^{\text{T}}
\end{array} & (\mathbf{X}\mathbf{B}_2\mathbf{D}_K + \mathbf{N}\mathbf{B}_K) \mathbf{C}_2
\end{bmatrix}\\[0.7cm]
&=&\begin{bmatrix}
\mathbf{B}_2\widehat{C}& \mathbf{B}_2 \mathbf{D}_K \mathbf{C}_2\\ 
\widehat{A}_{\mathscr{T}} & \widehat{B} \mathbf{C}_2
\end{bmatrix},\\[0.4cm] 
\mathscr{B} & = & \begin{bmatrix}
\mathbf{B}_1\\ 
\mathbf{X}\mathbf{B}_1
\end{bmatrix}, \mathscr{C} = \begin{bmatrix}
\mathbf{C}_1 \mathbf{Y}& \mathbf{C}_1 
\end{bmatrix}, \mathscr{C}_{\tau}  =  \begin{bmatrix}
\mathbf{C}_{1\tau} \mathbf{Y}& \mathbf{C}_{1\tau} 
\end{bmatrix},\\[0.1cm]
\mathscr{C}_{\mathscr{T}} & = & \begin{bmatrix}
\mathbf{D}_{12}(\mathbf{D}_K\mathbf{C}_2\mathbf{Y} + \mathbf{C}_K \mathbf{M}^{\text{T}}) & \mathbf{D}_{12} \mathbf{D}_K \mathbf{C}_2 
\end{bmatrix}\\[0.2cm]
& = & \begin{bmatrix}
\mathbf{D}_{12}\widehat{C} & \mathbf{D}_{12} \mathbf{D}_K \mathbf{C}_2 
\end{bmatrix},\\[0.1cm]
\widetilde{\boldsymbol{\Xi}}_{11}  & = &  \bigg[ \sum_{i=1}^s \pm \Big(\nu_i \frac{\partial \widetilde{\mathbf{P}}(\boldsymbol{\rho})}{\partial \rho_i}\Big) \bigg] + \widetilde{\mathbf{Q}}_\tau - \widetilde{\mathbf{R}}_{\tau} + \overline{\tau}^2 \widetilde{\mathbf{T}}_\tau + \widetilde{\mathbf{Q}}_{\mathscr{T}}  \\[0.2cm]
& - &  \widetilde{\mathbf{R}}_{\mathscr{T}} + \mathscr{A} +  \mathscr{A}^{\text{T}},
\end{array}
\end{equation*}
\end{small}
\begin{small}
\begin{equation*}
\begin{array}{lll}
\widetilde{\boldsymbol{\Xi}}_{33}  & = &  -\bigg[1- \sum_{i=1}^s \pm \Big(\nu_i \frac{\partial \tau}{\partial \rho_i}\Big) \bigg] \widetilde{\mathbf{Q}}_\tau -\widetilde{\mathbf{R}}_{\tau} + \lambda_3 (\mathscr{A}_\tau + \mathscr{A}^{\text{T}}_\tau),\\[0.1cm]
\widetilde{\boldsymbol{\Xi}}_{44}  & = &  -\bigg[1- \sum_{i=1}^s \pm \Big(\nu_i \frac{\partial \mathscr{T}}{\partial \rho_i}\Big) \bigg] \widetilde{\mathbf{Q}}_{\mathscr{T}} -\widetilde{\mathbf{R}}_{\mathscr{T}} + \lambda_4 (\mathscr{A}_{\mathscr{T}} + \mathscr{A}^{\text{T}}_{\mathscr{T}}).
\end{array}
\label{Eqcoef1}
\end{equation*}
\end{small}
\end{theorem}
\subsection{Output-Feedback LPV Controller Synthesis}
\noindent After obtaining the LMI decision variables, $\mathbf{X}$, $\mathbf{Y}$, $\widehat{A}$, $\widehat{A}_{\tau}$, $\widehat{A}_{\mathscr{T}}$, $\widehat{B}$, $\widehat{C}$, and $\mathbf{D}_K$ that satisfy (\ref{eq:LMImain}), the matrices of the continuous-time delayed LPV controller (\ref{eq:contcont1}) can be computed as follows:
\begin{small}
\begin{equation}
\begin{array}{rcl}
\mathbf{A}_{K}&=&\mathbf{N}^{-1} (\widehat{A} -  \mathbf{X} \mathbf{A} \mathbf{Y}) \mathbf{M} ^{\text{T}}, \\[0.10cm]
\mathbf{A}_{\tau K}&=&\mathbf{N}^{-1} (\widehat{A}_{\tau} -  \mathbf{X} \mathbf{A}_{\tau} \mathbf{Y}) \mathbf{M} ^{\text{T}}, \\[0.10cm]
\mathbf{B}_K&=&\mathbf{N}^{-1} (\widehat{B} -  \mathbf{X} \mathbf{B}_2 \mathbf{D}_K),\\[0.10cm]
\mathbf{C}_K&=&(\widehat{C} - \mathbf{D}_K \mathbf{C}_{2} \mathbf{Y}) \mathbf{M} ^{-\text{T}},\\[0.10cm]
\mathbf{A}_{\mathscr{T} K}&=&\mathbf{N}^{-1} ( \widehat{A}_{\mathscr{T}} - \mathbf{X}\mathbf{B}_2\mathbf{D}_K\mathbf{C}_2\mathbf{Y} - \mathbf{N}\mathbf{B}_K\mathbf{C}_2\mathbf{Y} \\[0.10cm]
&-& \mathbf{X} \mathbf{B}_2 \mathbf{C}_{K} \mathbf{M} ^{\text{T}}) \mathbf{M} ^{-\text{T}},
\end{array}
\label{eq:cont_matrices}
\end{equation}
\end{small}
\noindent where $\mathbf{M}$ and $\mathbf{N}$ are square and invertible matrices and are obtained from the factorization problem
\begin{equation}
\mathbf{I} - \mathbf{X}\mathbf{Y} = \mathbf{N} \mathbf{M}^{\text{T}}.
\end{equation} 

\subsection{Digital Controller Derivation}

For implementation purposes, we need to find a discretized equivalence of the designed continuous-time LPV control design (\ref{eq:contcont1}) as follows:
\begin{equation}
\begin{array}{rcl}
\mathbf{x}_d(k+1) & = & \mathbf{A}_d (\boldsymbol{\rho}(k)) \mathbf{x}_d(k)+ \sum_{i=1}^{l+2} \mathbf{A}_{\tau d_i} (\boldsymbol{\rho}(k)) \mathbf{x}_d(k-i)\\[0.15cm]
 & + & \mathbf{B}_d(\boldsymbol{\rho}(k))\mathbf{y}(k), \\[0.15cm]
\mathbf{u}_d(k) & = & \mathbf{C}_d (\boldsymbol{\rho}(k)) \mathbf{x}_d(k)+ \mathbf{D}_d(\boldsymbol{\rho}(k))\mathbf{y}(k).
\end{array}
\label{eq:discretecontroller2}
\end{equation}
\noindent By using the approached suggested by \cite{ramezanifar2014output}, the discrete-time controller matrices are computed as follows:
\begin{equation}
\begin{array}{rcl}
\mathbf{A}_d & = & e^{(t_{k+1}-t_k) \mathbf{A}_K(\boldsymbol{\rho}(t_k))}+(e^{(t_{k+1}-t_k) \mathbf{A}_K(\boldsymbol{\rho}(t_k))}-\mathbf{I}) \\[0.15cm]
&\times&  \mathbf{A}_K^{-1}(\boldsymbol{\rho}(t_k)) \mathbf{A}_{\mathscr{T} K}(\boldsymbol{\rho}(t_k)), \\[0.2cm]
\mathbf{B}_d &=& (e^{(t_{k+1}-t_k) \mathbf{A}_K(\boldsymbol{\rho}(t_k))}-\mathbf{I})\mathbf{A}_K^{-1}(\boldsymbol{\rho}(t_k)) \mathbf{B}_{K}(\boldsymbol{\rho}(t_k)),\\[0.10cm]
\mathbf{C}_d & = & \mathbf{C}_{K}(\boldsymbol{\rho}(t_k)) , \\[0.10cm]
\mathbf{D}_d & = & \mathbf{D}_{K}(\boldsymbol{\rho}(t_k)) , \\[0.15cm]
\mathbf{A}_l&\!=\!&\! \dfrac{c_1}{2}( e^{(t_{k+1}-t_k) \mathbf{A}_K(\boldsymbol{\rho}(t_k))} \!-\! e^{(t_{k+1}-\tau_k -t_{l+1}) \mathbf{A}_K(\boldsymbol{\rho}(t_k))} )\\[0.10cm]
&\times&  \mathbf{A}_K^{-1}(\boldsymbol{\rho}(t_k)) \mathbf{A}_{\tau K}(\boldsymbol{\rho}(t_k)), \\[0.2cm]
\mathbf{A}_{l+1}&=& \big(\dfrac{1+c_2}{2} e^{(t_{k+1}-t_k) \mathbf{A}_K(\boldsymbol{\rho}(t_k))}\\[0.15cm]
& - &  \dfrac{c_2 - c_3}{2} e^{(t_{k+1}-\tau_k -t_{l+1}) \mathbf{A}_K(\boldsymbol{\rho}(t_k))} - \dfrac{1+c_3}{2} \mathbf{I}\big) \\[0.15cm]
& \times & \mathbf{A}_K^{-1}(\boldsymbol{\rho}(t_k)) \mathbf{A}_{\tau K}(\boldsymbol{\rho}(t_k)), \\[0.15cm]
\mathbf{A}_{l+2}&=& \dfrac{c_4}{2} (e^{(t_{k+1}-\tau_k -t_{l+1}) \mathbf{A}_K(\boldsymbol{\rho}(t_k))} - \mathbf{I}) \\[0.15cm]
& \times &   \mathbf{A}_K^{-1}(\boldsymbol{\rho}(t_k)) \mathbf{A}_{\tau K}(\boldsymbol{\rho}(t_k)),
\end{array}
\label{eq:cont_matrices}
\end{equation}
\noindent with 
\begin{equation*}
\begin{array}{rcl}
c_1 & = & \dfrac{t_{l+1}-(t_k - \tau_k)}{t_{l+1}-t_l},
c_2  =  \dfrac{(t_k - \tau_k)-t_l}{t_{l+1}-t_l},\\[0.20cm]
c_3 & = & \dfrac{t_{l+2}-(t_{k+1} - \tau_k)}{t_{l+2}-t_{l+1}},
c_4  =  \dfrac{(t_{k+1} - \tau_k)-t_{l+1}}{t_{l+2}-t_{l+1}}.
\end{array}
\end{equation*}

\section{AFR Sampled-Data Control Design Simulation Results and Discussions}\label{sec:Results}

\begin{figure} 
\centering \includegraphics[width=0.9\columnwidth,height=1.4in]{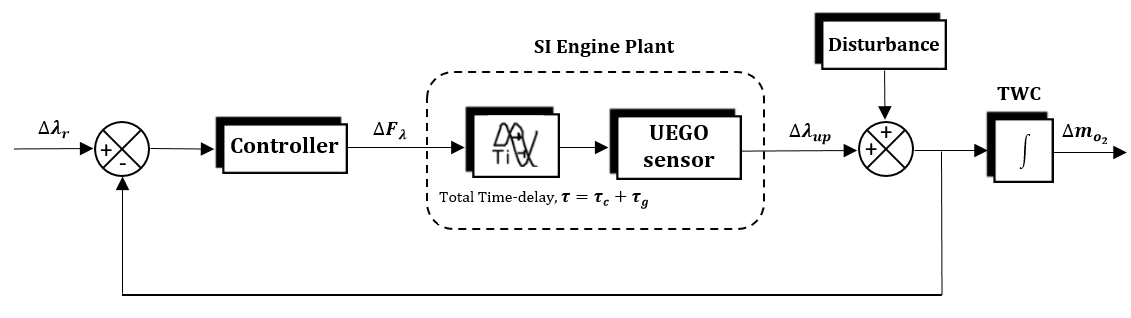} 
\caption{AFR closed-loop system structure} 
\label{fig:closed-loop system structure}
\end{figure}

In this section, we assess the performance of the proposed delay-dependent sampled-data LPV control design approach in the AFR reference profile tracking in SI engines under various simulated scenarios. Figure \ref{fig:closed-loop system structure} demonstrates the structure of the closed-loop system.

The dynamic properties of the studied AFR control problem follows (\ref{engine_final_LPV}). By considering the state-space representation of a general LPV time-delay system (\ref{eq:LPVsystem}), the controlled output vector, $\mathbf{z}(t)$, in (\ref{eq:LPVsystem}), is defined as $\mathbf{z}(t) = [\phi \cdot x_3 (t), \:\: \psi \cdot x_4 (t),  \:\: \xi \cdot u(t)]^{\text{T}}$, \textit{i.e.} \begin{small}
$\mathbf{C}_{1}(\boldsymbol{\rho}(t))=\begin{bmatrix}
0 & 0 & \phi & 0\\
0 & 0 & 0 & \psi \\
0 & 0 & 0 & 0
\end{bmatrix}$,
\end{small} $\mathbf{D}_{12}(\boldsymbol{\rho}(t))=[0 \:\: 0 \:\: \xi]^{\text{T}}$, and $\mathbf{D}_{11} = \mathbf{0}_{3\times2}$. The command tracking performance, TWC oxygen storage behavior,  and the control effort are being penalized by the weighting scalars $\phi$, $\psi$, and $\xi$, respectively to fulfill desired performance objectives. 


\begin{remark}
The conditions provided in Theorems \ref{thm:thm1} result in an infinite-dimensional convex optimization problem with an infinite number of LMI constraints. To overcome this challenge and compute a solution, we utilize the gridding method to approximate the infinite-dimensional problem with a finite-dimensional convex optimization one \cite{apkarian1998advanced}. Additionally, affine parameter dependence has been adopted as follows: $\mathbf{M}(\boldsymbol{\rho}(t))=\mathbf{M}_0 + \sum\limits_{i=1}^{n_s}\rho_i(t) \mathbf{M}_{i_1}$, where $\mathbf{M}(\boldsymbol{\rho}(t))$ represents any of the involved parameter-dependent LMI decision matrix variables. Finally, gridding the scheduling parameter space at proper intervals leads to a finite set of LMIs to be solved for the unknown matrices and $\gamma$. Also, in order to improve the results, a $3$-dimensional search involving the three scalar variables $\lambda_2$, $\lambda_3$, and $\lambda_4$ is performed to obtain the minimum value of the performance index, $\gamma$. The MATLAB\textsuperscript{\tiny\textregistered} toolbox YALMIP with Mosek solver is used to solve the corresponding LMI optimization problems \cite{lofberg2004yalmip}.
\end{remark}

\begin{figure}[t] 
\centering \includegraphics[width=3.2in,height=1.6in]{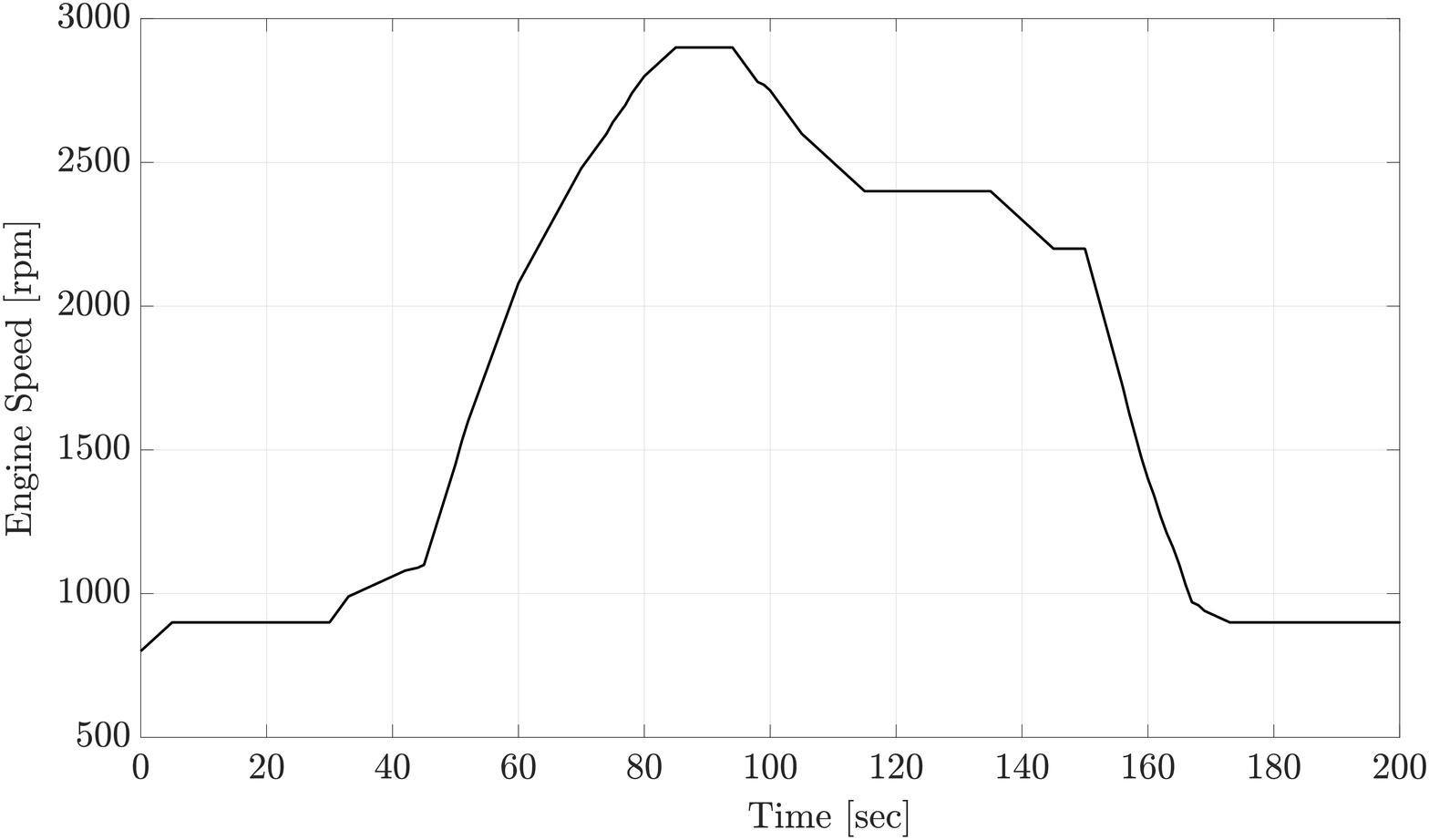} 
\caption{Engine operation speed range} 
\label{fig:Engine speed}
\end{figure}

For the considered simulation scenarios, the engine speed (LPV scheduling parameter) variation is assumed to be between $800$ rpm (idle) and $4000$ rpm (high speed), \textit{i.e.} $\omega(t) \in [800, \:\: 4000]$. Figure \ref{fig:Engine speed} shows the speed range, over which the engine operates, that captures the effects of running idle, low speed, acceleration, and braking events, respectively. In order to have an effective sampled-data control design, the choice of sampling frequency is crucial. There is a trade-off between the quality of the closed-loop system response and the implementation cost. In the engine problem, the sampling rate depends on the engine speed (rpm). As a result, the sampling period is also not fixed and varies accordingly. In this study, we consider one sample per two revolutions of the engine, \textit{i.e.}
\begin{equation}
t_{k+1} = t_k + \dfrac{4\pi}{\omega(t_k)}.
\end{equation}

For comparison purposes, we considered a delay-free output-feedback LPV control design, implemented in \cite{zope2009air}. In \cite{zope2009air} a controller for the continuous-time delay-free LPV model of AFR dynamics is designed, which has been obtained using the Pad\'{e} approximation to approximate the LPV time-delay system to a delay-free one. Additionally, we utilize the trapezoidal approximation \cite{apkarian1997discretization} to discretize the continuous-time LPV controller, and finally, end up with an indirect conventionally designed digital controller ignoring the sampling/hold rate implicitly in the control design process.

\begin{figure}[t] 
\centering \includegraphics[width=\columnwidth,height=1.95in]{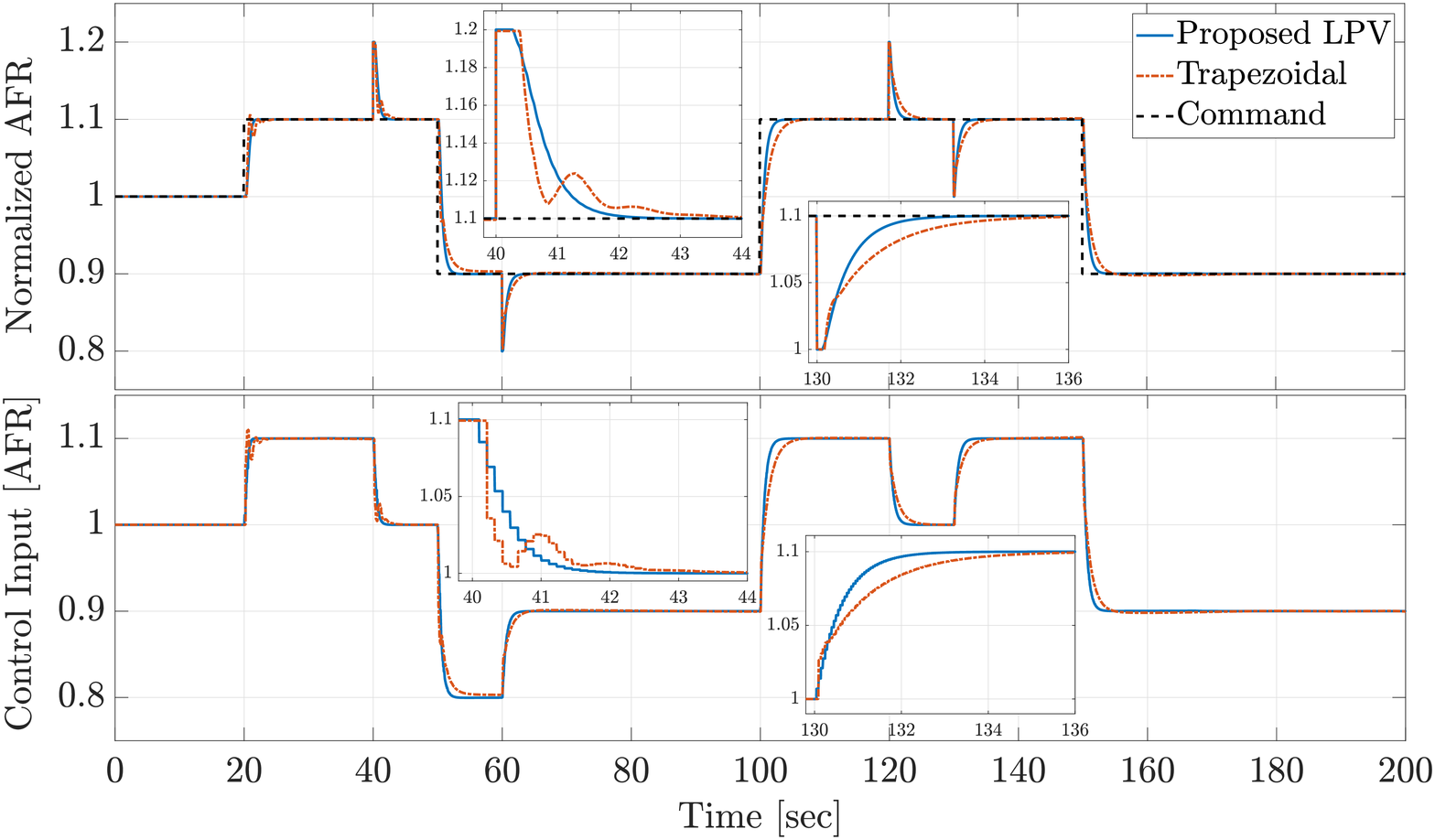} 
\caption{AFR tracking performance and control efforts for disturbance-free case} 
\label{fig:tracking w/o disturbance}
\end{figure}

\begin{figure}[t] 
\centering \includegraphics[width=\columnwidth,height=1.95in]{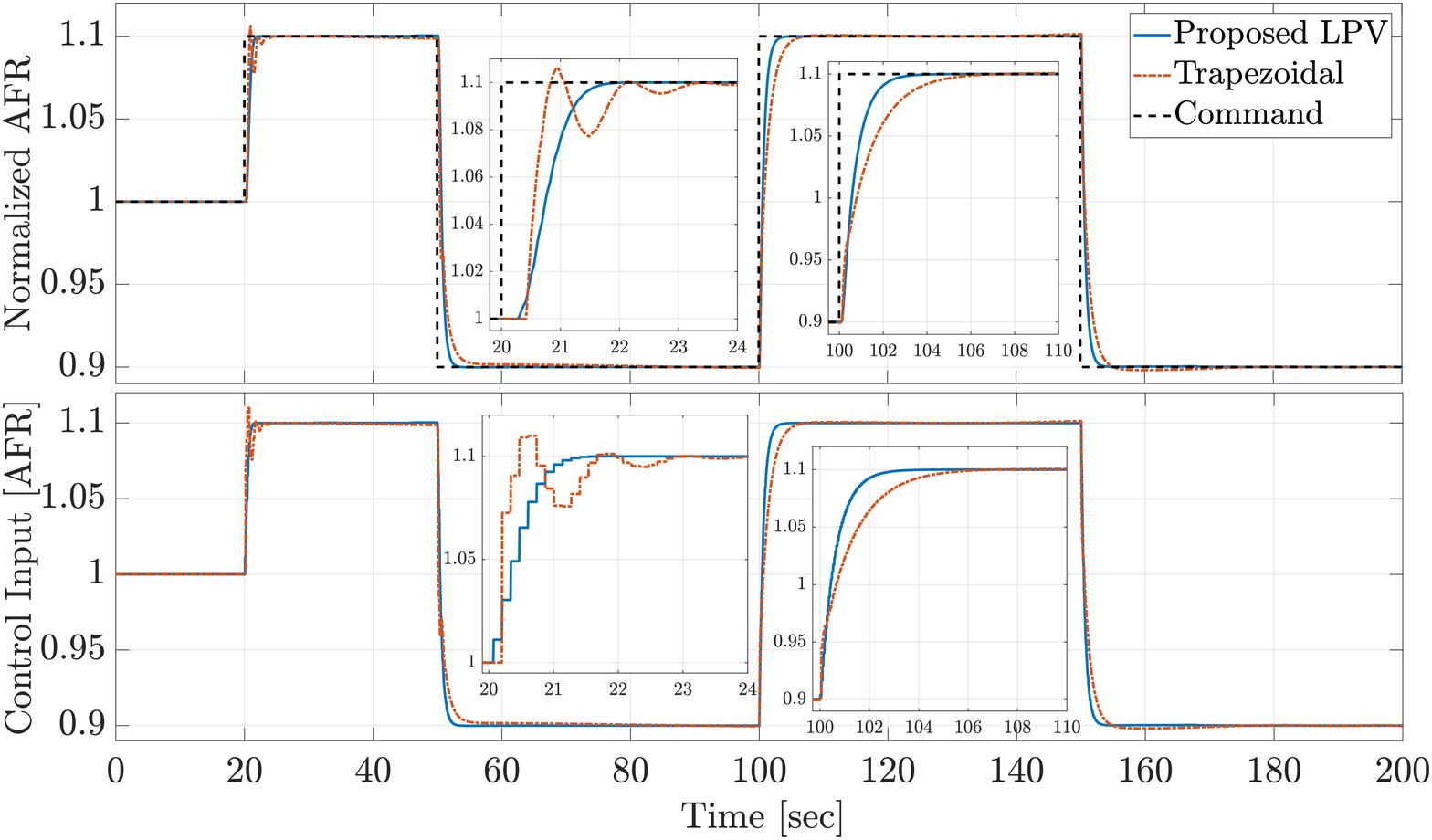} 
\caption{AFR tracking performance and control efforts under output disturbance} 
\label{fig:overall tracking}
\end{figure}

\begin{figure}[t] 
\centering \includegraphics[width=0.80\columnwidth,height=1.37in]{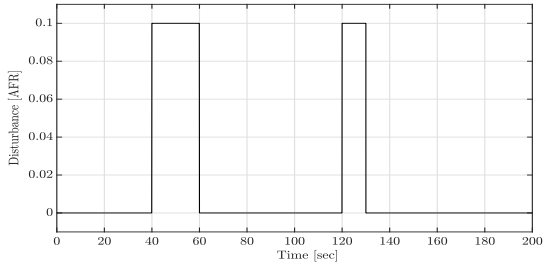} 
\caption{Disturbance profile} 
\label{fig:Typical disturbance profile}
\end{figure}

The first simulation scenario has been created to evaluate the AFR reference tracking performance of the controller in the absence of any disturbance. The tracking profiles and the step-wise discrete control efforts of both the proposed output-feedback sampled-data controller and the LPV controller taken from \cite{zope2009air} are shown in Fig. \ref{fig:tracking w/o disturbance} where the objective is to track the commanded AFR with the minimum overshoot and settling time and zero steady-state error. The AFR command profile is initially set to unity until time $t = 20 \: sec$, and then, it alters between lean and rich regions, namely $\lambda_r=1.1$ and $0.9$.  It is noted that both the time delay and the sampling interval are time-varying as they are functions of the engine speed. Figure \ref{fig:overall tracking} depicts the performance of both controllers, where the closed-loop system is subject to output disturbances. These external output disturbances act on the fuel-path of the SI engine and are caused by perturbations in the cylinder air-flow induced by the tip-ins and tip-outs, the periodic purge cycles, and the model mismatches. Such a disturbance profile is shown in Fig. \ref{fig:Typical disturbance profile}. As Figs. \ref{fig:tracking w/o disturbance} and \ref{fig:overall tracking} demonstrate, the proposed time-delayed LPV gain-scheduling controller outperforms the conventional indirect control design in terms of reference command tracking and disturbance rejection.

\begin{figure}[t]
    \centering
    \subfigure[]
    {
        \includegraphics[width=0.98\columnwidth, height=1.85in]{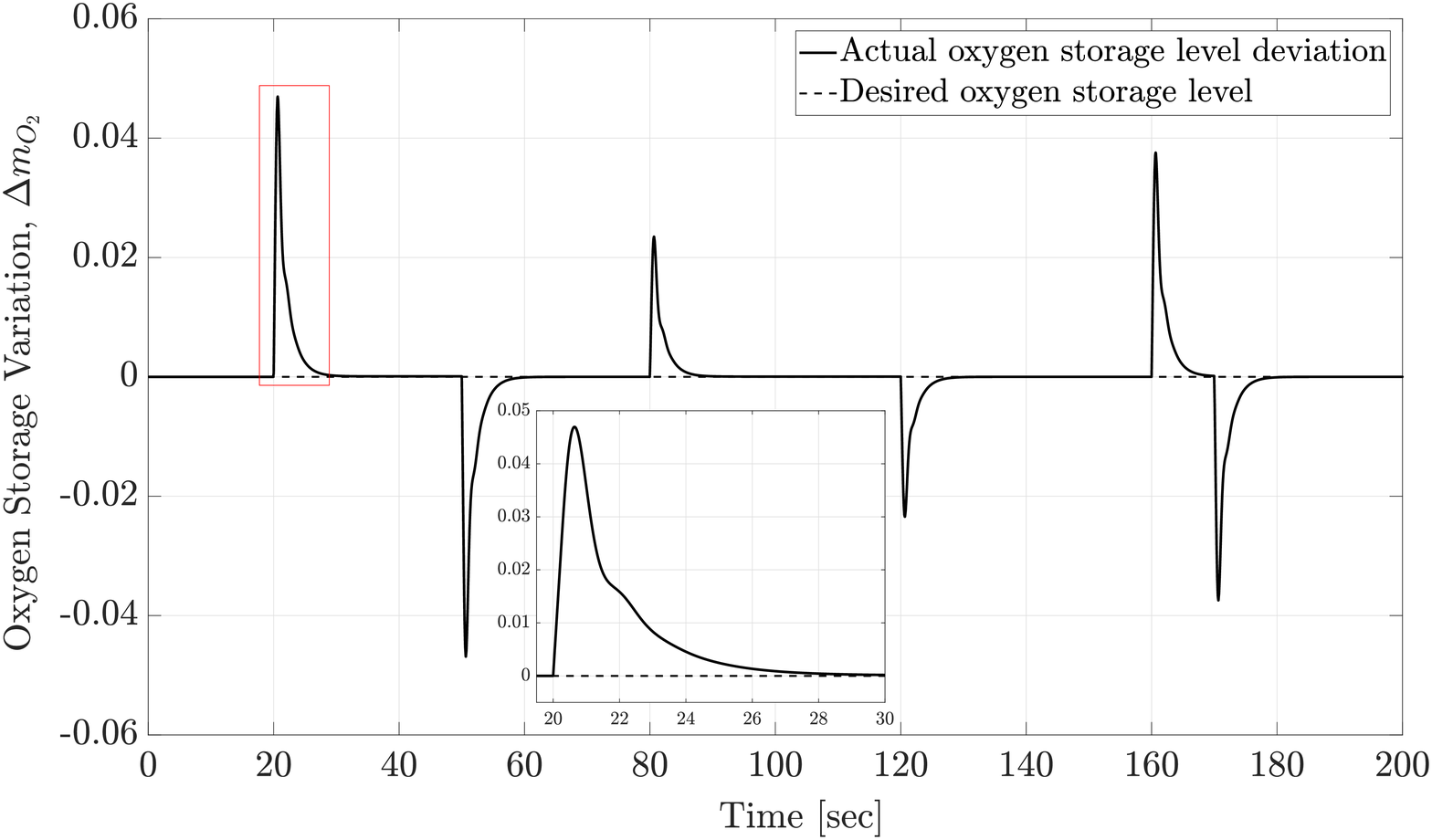}
\label{fig:oxygen storage variation800}
    }
    \subfigure[]
    {
        \includegraphics[width=0.98\columnwidth, height=1.85in]{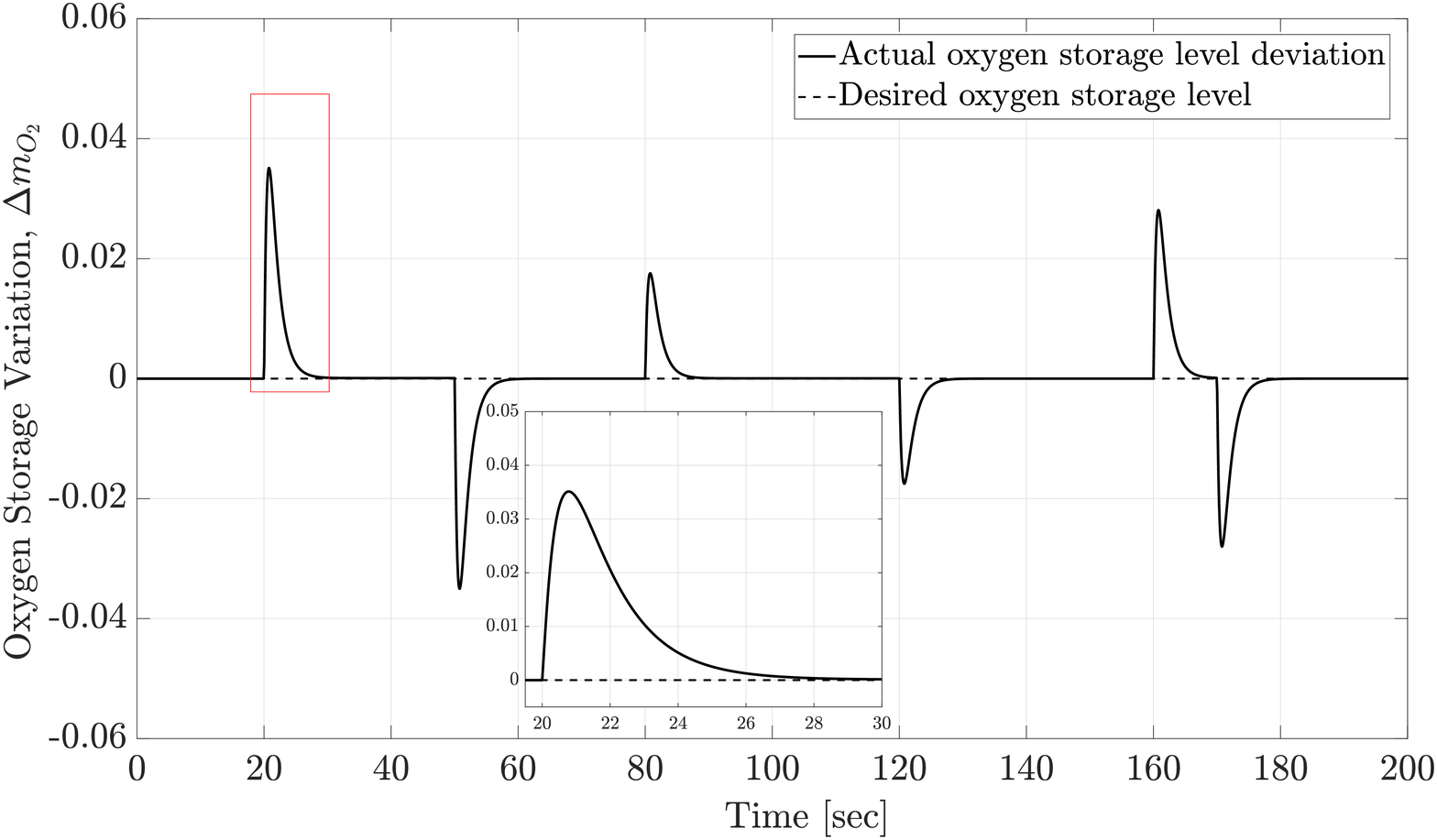}

\label{fig:oxygen storage variation3000}
    }
\caption{Oxygen storage variation $\Delta m _{O_2}$ in TWC in the presence of the external disturbance for engine operating speeds (a) $800$ rpm (b) $3000$ rpm} 
    \label{fig:uncertainty}
\end{figure}

\begin{figure}[t] 
\centering \includegraphics[width=0.80\columnwidth,height=1.37in]{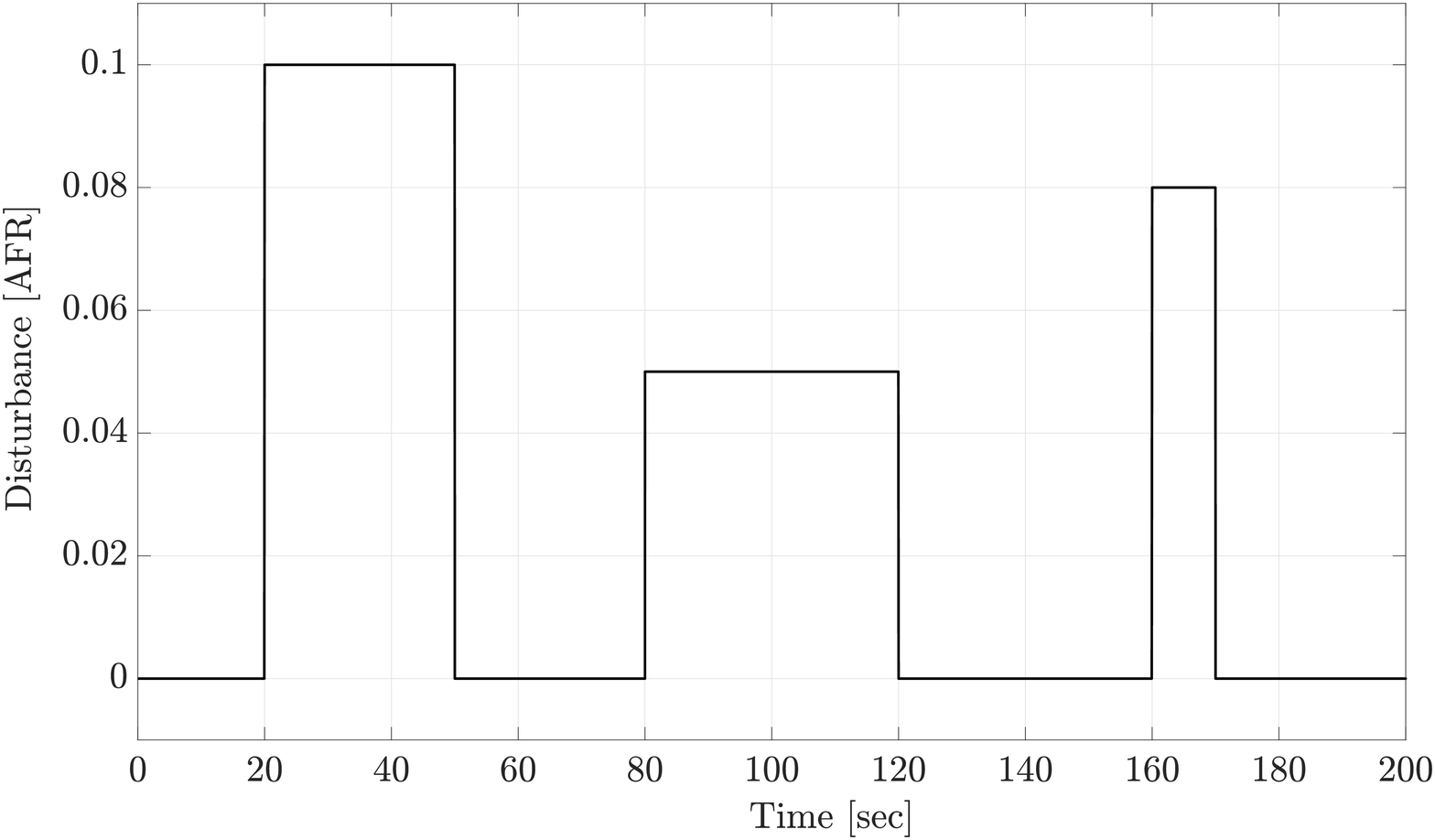} 
\caption{Disturbance profile for the oxygen storage variation} 
\label{fig:disturbance profile2}
\end{figure}

Next, in order to validate the effectiveness of the proposed control design approach in maintaining the mass of stored oxygen, $m_{O_2}$, in the TWC close to its desired value of 50$\%$, we set the reference command to be $\lambda_r = 1$ (stoichiometric AFR). Figures \ref{fig:oxygen storage variation800} and \ref{fig:oxygen storage variation3000} depict the deviation of the mass of stored oxygen ($m_{O_2}$) at a low engine speed ($800$ rpm), which corresponds to larger time-delay, and at a high speed engine ($3000$ rpm), where the time-delay is comparatively smaller. To this aim, a new external disturbance profile is considered, as shown in Fig. \ref{fig:disturbance profile2}. Figures \ref{fig:oxygen storage variation800} and \ref{fig:oxygen storage variation3000} confirm that the disturbance effect has been rejected and the deviation in the level of oxygen stored in the TWC has returned back to the demanded value of 50$\%$ (zero variation in $\Delta m_{O_2}$) quickly. The simulations demonstrate that the proposed sampled-date LPV control design scheme has an appreciable utility for sampled-data LPV time-delay systems as evident from the benchmark problem of AFR control in an SI engine examined in this paper. Specifically, the sampled-data LPV control design, due to its scheduling structure and direct discrete-time synthesis which takes the sampling/hold times into account, has demonstrated a superior AFR tracking performance with respect to the rise time and speed of the response while desirably rejecting the external disturbances over the entire range of the engine operation.

\section{Conclusion}
\label{sec:Conclusion}
Accurate AFR regulation is essential to improve fuel economy and to attain the high efficiency of the TWC and consequently minimize tailpipe emissions in SI engines. The position of the oxygen sensor introduces a varying time-delay in the feedback loop. In this regard, the speed-dependent dynamics of the fuel-path of the SI engine has been modeled as an LPV time-delay system. Due to the implementation of the digital controller, the closed-loop system is a hybrid one, so, the input-delay technique has been utilized to convert the hybrid closed-loop system into a continuous-time state-delay LPV system, which is suitable for sampled-data output-feedback control design.  To this end, the designed sampled-data LPV has been addressed to track the setpoint AFR profile in the face of a varying time delay, varying sampling time, and external disturbances. Stabilization criteria are obtained via bounding the derivative of a Lyapunov-Krasovskii functional candidate, and the results are formulated in a parameter-dependent LMI setting. By means of simulations, it is shown that the proposed controller is capable of meeting the AFR performance requirements and outperforms conventional controllers. Future research will aim to address the sampled-data LPV control problem in the presence of parameter and time-delay uncertainties.


\end{document}